\documentclass[aps,pre,twocolumn,groupedaddress,showpacs,10 pt]{revtex4}
\usepackage{graphicx}
\usepackage{amsmath}
\usepackage{amssymb,color}
\usepackage[colorlinks,linkcolor=blue,citecolor=blue]{hyperref}
\usepackage{ bbold }
\usepackage{ wasysym }

\begin{document}
\title{Suppression of deterministic and stochastic extreme desynchronization events using anticipated synchronization}
\author{Jordi Zamora-Munt, Claudio R. Mirasso, and Raul Toral}
\affiliation{Instituto de F\'isica Interdisciplinar y Sistemas Complejos, IFISC (CSIC-UIB), E-07122 Palma de Mallorca - Spain}

\date{\today}

\begin{abstract}
We numerically show that extreme events induced by parameter mismatches or noise in coupled oscillatory systems can be anticipated and suppressed before they actually occur. We show this in a main system unidirectionally coupled to an auxiliary system subject to a negative delayed feedback. Each system consists of two electronic oscillators coupled in a master-slave configuration. Extreme events are observed in this coupled system as large and sporadic desynchronization events. Under certain conditions, the auxiliary system can predict the dynamics of the main system. We use this to efficiently suppress the extreme events by applying a direct corrective reset to the main system.
\end{abstract}
\maketitle

Extreme events can be defined as those whose amplitude or duration, being much larger than the average, are representative of the tail of a probability distribution \cite{ExtremeBook}. This kind of events has been reported in different disciplines ranging from climatology to optics, population dynamics or economy \cite{pre_2009,cc_2010,nature_2007,ol_2013,epj_2012}, and its study has attracted the interest of researchers in recent years due to their possible destructive effects. Many efforts have been devoted to understand their origin, predict when and where they will appear and, if possible, suppress them \cite{Nicolis06,Sornette09, Sornette12}.

Some of these extreme events are irregular and sporadic, and considered unpredictable: they appear spontaneously without a precursor and any clue on how they will evolve \cite{RogueBook}. However, recent studies have shown that some extreme events can be deterministic \cite{Bonatto11, Zamora13} suggesting they can be predicted. Meanwhile, it is not clear if truly random extreme events (induced by some source of noise) can also be predicted and suppressed, still constituting a challenge in current research.

Extreme stochastic or deterministic events can be found in coupled systems as sporadic desynchronization intervals induced by noise or parameter mismatches. These desynchronization events are also known as ``bubbling'' and have been observed and intensively studied, for instance, in semiconductor lasers \cite{Flunkert09,Tiana12} and electronic circuits \cite{Gauthier96}. Recently, Cavalcante and coworkers \cite{Cavalcante13} studied bubbling events of two unidirectionally coupled electronic circuits and observed that the largest events are more frequent than what one could expect extrapolating the power-law distribution of the small ones. This property characterizes a particular type of extreme events known as Dragon Kings (DKs) \cite{Sornette09}.

A plausible method to predict DKs is by anticipating the events. This could be done through the anticipated synchronization that can be achieved by unidirectionally coupling an auxiliary system to a main system \cite{Voss00}. Under certain conditions the dynamics of the auxiliary system is identical to that of the main one but advanced a certain time. It has been demonstrated that this scheme efficiently predicts chaotic and excitable dynamics \cite{Masoller01,Sivaprakasam01,Ciszak03}. Furthermore, it has been used to suppress noise-induced spikes \cite{Ciszak09,Mayol12} applying an appropriate external input to the target system.

The aim of this paper is to elucidate if it is possible to anticipate extreme desynchronization events and suppress them by using anticipated synchronization. The proposed method allows us to know in advance the dynamics of a target system and apply corrective resets to keep these events under a safety amplitude. In section \ref{model} we describe the system used to obtain the extreme events, the coupling scheme and parameters that will lead to stable anticipated synchronization. In section \ref{prevention} we study the cases of deterministic events (\ref{deterministic}) induced by parameter mismatch and stochastic events induced by and external noise source (\ref{stochastic}). Finally, in section \ref{conclusions} we summarize our main results.

\section{Prediction of extreme events}
\label{model}

The coupling scheme we use to obtain anticipated synchronization is given by a system of two unidirectionally delay-coupled oscillators as introduced in \cite{Voss00},
\begin{eqnarray}
\dot{\mathbf{x}}&=&f(\mathbf{x}(t)), \label{eq:x}\\
\dot{\mathbf{y}}&=&f(\mathbf{y}(t))+\mathbf{C}(\mathbf{x}(t)-\mathbf{y}(t-\tau)),\label{eq:y}
\end{eqnarray}
where $\mathbf{x}$ is the main system and $\mathbf{y}$ is the auxiliary system. The dot represents time derivatives, $\mathbf{C}$ is the matrix of coupling strengths, and $\tau$ is the feedback delay time. In some parameter regions the solution $\mathbf{x}(t+\tau)=\mathbf{y}(t)$ is stable, such that the system $\mathbf{y}$ predicts the behavior of $\mathbf{x}$ a time $\tau$ in advance.

\begin{figure}[]
\includegraphics[width=8cm] {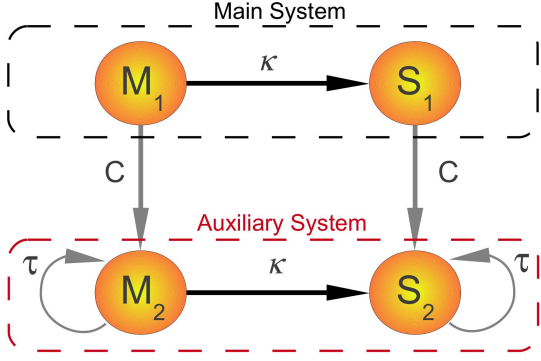}
\caption{\label{fig:1} Coupling scheme used for anticipated synchronization. Each subsystem consist of two unidirectionally coupled oscillators in a master-slave configuration. The main system is unidirectionally coupled to the auxiliary system according to eqs. (\ref{eq:x})-(\ref{eq:y}).}
\end{figure}

\begin{figure}[]
\includegraphics[width=8cm] {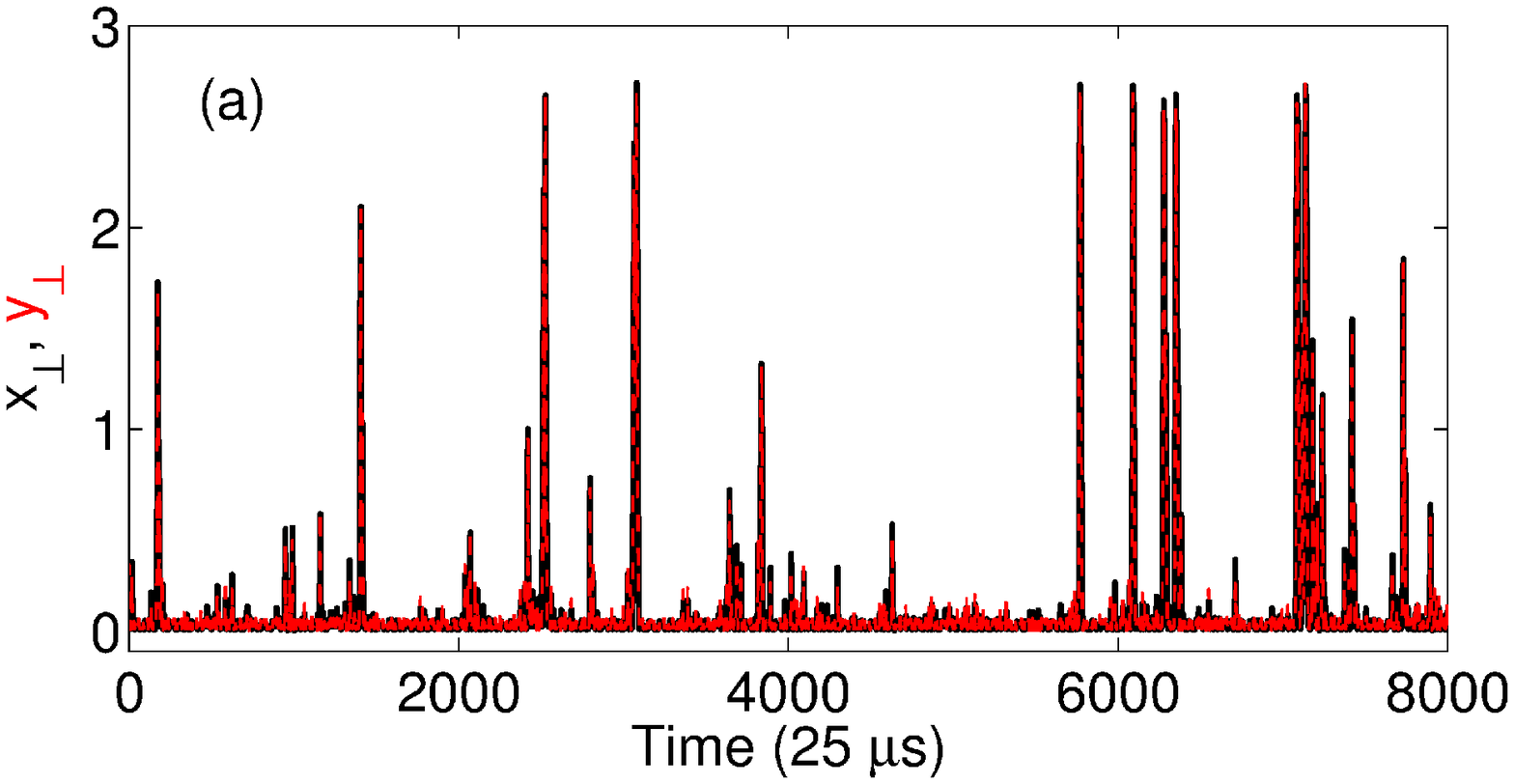}
\includegraphics[width=8cm] {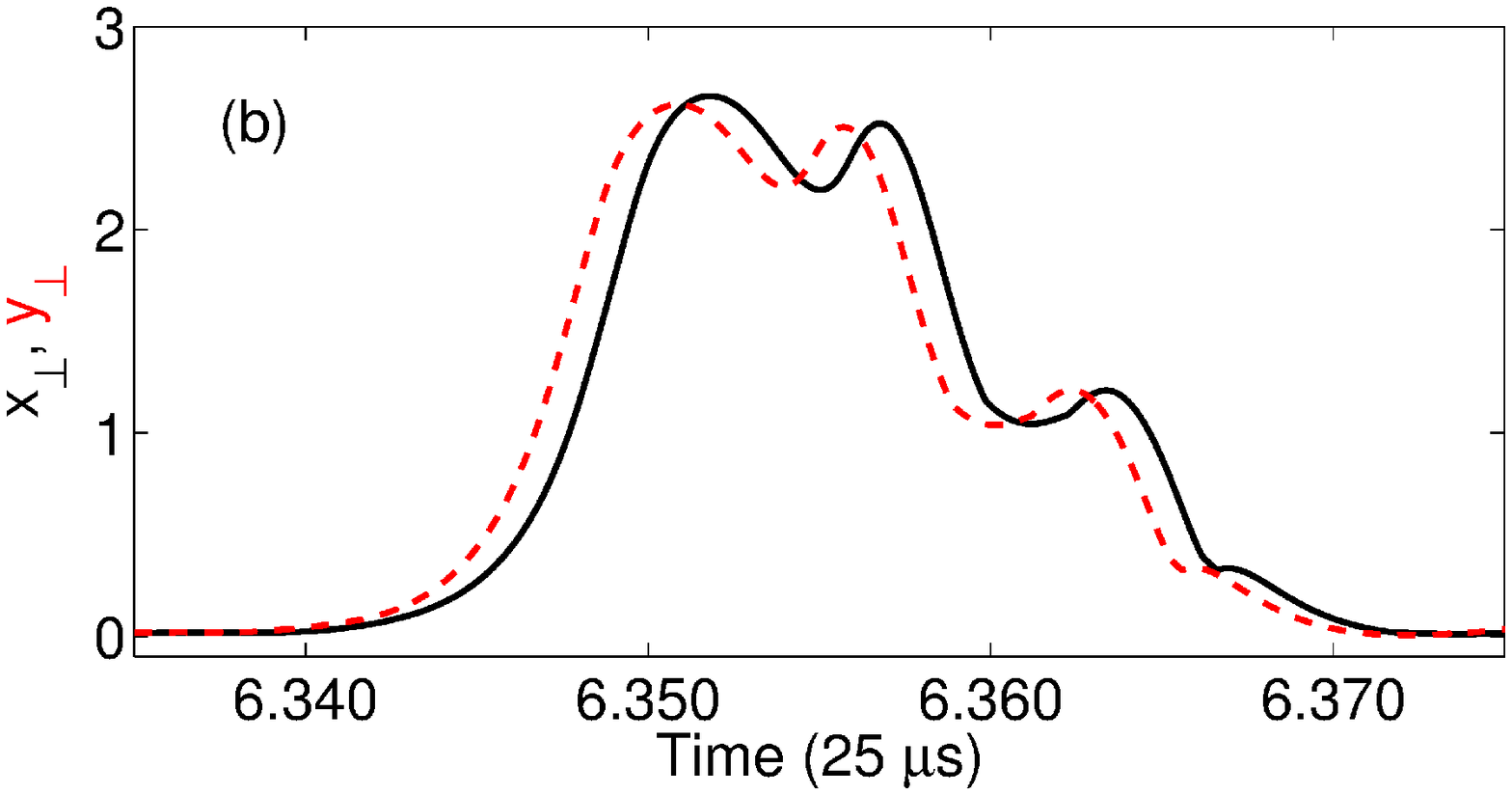}
\caption{\label{fig:3} (Color online) (a) Time trace of $x_{\perp}$ for the main system (solid line) and $y_{\perp}$ for the predicted events by the auxiliary system (dashed line). (b) A magnification of an extreme event in the main system (solid line) and the predicted events by the auxiliary system (dashed line). $\tau=1$ and $C=$0.75 and the other parameters are listed in Table \ref{tab:T1}.}
\end{figure}

We are interested in predicting desynchronization events that occur in two unidirectionally coupled electronic circuits in a master-slave configuration \cite{Cavalcante13}. This system plays the role of the main system $\mathbf{x}=(V,v,I)$ as shown in figure \ref{fig:1} and its dynamical evolution is given by
\begin{eqnarray}
\dot{V}_{j_1}&=&\frac{V_{j_1}}{R_{1j_1}}-g_{j_1}\left[V_{j_1}-v_{j_1}\right], \label{eq:V1}\\
\dot{v}_{j_1}&=&g_{j_1}\left[V_{j_1}-v_{j_1}\right]-I_{j_1}, \label{eq:v1}\\
\dot{I}_{j_1}&=&v_{j_1}-R_{4j_1}I_{j_1}, \label{eq:I1}\end{eqnarray}
where
\begin{eqnarray}
g_{j}\left[V\right]&=&\frac{V}{R_{2,j}}+I_{r,j}\left(e^{\alpha_{f,j}V}-e^{-\alpha_{r,j}V}\right). \label{eq:g} \end{eqnarray}
The subscript $j_1$ can be ``$M_1$'' for the master or ``$S_1$'' for the slave. The coupling term from the master to the slave is added in equation (\ref{eq:v1}) for the slave as $\kappa\left(v_{M_1}-v_{S_1}\right)$ where $\kappa$ is the coupling strength. We arbitrarily coupled through $v_{S_1}$ because synchronization is less stable than through $V_{S_1}$ \cite{Gauthier96}. The values of the different parameters used in the simulations are listed in Table \ref{tab:T1} where the time is normalized to $T=25\mu s$. Desynchronization events occur in this system due to a small mismatch between the parameters of the master and the slave or due to noise and can be visualized via the distance in the phase space (Fig. \ref{fig:3}, black solid line) as
\begin{eqnarray}
x_{\perp}=|V_{M_1}-V_{S_1}|+|v_{M_1}-v_{S_1}|+|I_{M_1}-I_{S_1}|.
\end{eqnarray}
The sizes of the small desynchronization events (roughly $x_{\perp}<2$) are distributed according to a power-law but the largest desynchronization events are more frequent than what one could expect from this distribution, which have been identified as DKs \cite{Cavalcante13}.

The auxiliary system is devoted to predict the dynamics of the main system and plays the role of $\mathbf{y}=(V,v,I)$ in eq. (\ref{eq:y}). Its time evolution is given by
\begin{eqnarray}
\dot{V}_{j_2}&=&\frac{V_{j_2}}{R_{1j_2}}-g_{j_2}\left[V_{j_2}-v_{j_2}\right] \nonumber \\
&&+C\left(V_{j_1}-V_{j_2}\left(t-\tau\right)\right),\label{eq:AuxV}\\
\dot{v}_{j_2}&=&g_{j_2}\left[V_{j_2}-v_{j_2}\right]-I_{j_2}, \label{eq:Auxv}\\
\dot{I}_{j_2}&=&v_{j_2}-R_{4j_2}I_{j_2},\label{eq:AuxI}
\end{eqnarray}
where the subscript $j_2$ refers to ``$M_2$'' for the master or ``$S_2$'' for the slave and the rest of parameters have the same meaning as in eqs. (\ref{eq:V1})-(\ref{eq:I1}). As was done for the main system, the coupling term $\kappa\left(v_{M_2}-v_{S_2}\right)$ is added in the equation (\ref{eq:Auxv}) for the slave. Additionally, the main system is coupled to the auxiliary system with a coupling strength given by $C$ and a delay time given by $\tau$, arbitrarily coupling them through $V$ because synchronization is more stable than through $v$ \cite{Gauthier96}. Following the notation of eq. (\ref{eq:y}), the distance between the master and the slave systems in the phase space is defined by
\begin{eqnarray}
y_{\perp}=|V_{M_2}-V_{S_2}|+|v_{M_2}-v_{S_2}|+|I_{M_2}-I_{S_2}|.
\end{eqnarray}

\begin{table} [htb]
\begin{ruledtabular}
\caption{\label{tab:T1} Parameters of the master and slave circuits in the main ($M_1$ and $S_1$) and auxiliary ($M_2$ and $S_2$) systems. They correspond to a parameter mismatch $\sim1-2\%$ with respect to $M_1$.}
\begin{tabular}{lcccc}
Parameter & $M_1$ & $S_1$ & $M_2$ & $S_2$\\
\hline
$I_{r}$ ($\mu A$) & 22.5  & 22.4 & 22.3 & 22.6 \\
$\alpha_{f}$ & 11.6 & 11.5 & 11.7 & 11.8 \\
$\alpha_{r}$ & 11.57 & 11.71 & 11.8 & 11.43 \\
$R_{1}$  & 1.298 & 1.308 & 1.32 & 1.28 \\
$R_{2}$ & 3.44 & 3.47 & 3.41 & 3.5 \\
$R_{4}$ & 0.193 & 0.195 & 0.191 & 0.2 \\
$\kappa$ & 0 & 4.6 & 0 & 4.6
\end{tabular}
\end{ruledtabular}
\end{table}

\begin{figure}[]
\includegraphics[width=8cm] {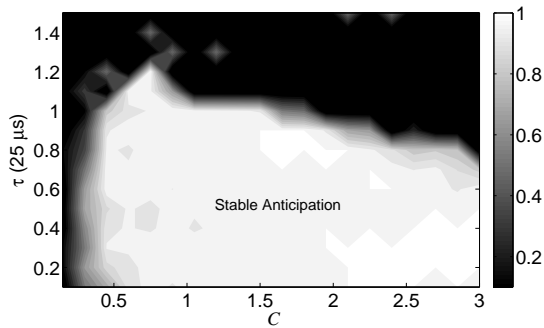}
\caption{\label{fig:4} Contour plot of the accuracy, $\rho$, in the plane ($C$, $\tau$).}
\end{figure}

Using the parameters given in Table \ref{tab:T1} the auxiliary system can predict accurately the extreme events as shown in figure \ref{fig:3} (red dashed line). Actually, larger events can be more accurately predicted than the smaller ones. Here we show a set of parameters but similar conclusions can be obtained with wider parameter mismatches. The accuracy, $\rho$, of the prediction is shown in figure \ref{fig:4} in the delay time vs. coupling strength parameter space. $\rho$ is calculated as the ratio of the number of maxima in the auxiliary divided by the number of maxima in the main system. Since we are interested in to predict the largest events, we only considered in the calculation of $\rho$ maxima larger than 0.3 times the absolute maximum in a realization and discarded the events smaller than this value. Stable anticipated synchronization occurs for $C>0.5$ and for $\tau\lesssim1$ and the anticipation time is on average given by the time delay $\tau$ \cite{footnote}. This result is consistent with 
results obtained in other system \cite{Voss00, Ciszak09, Mayol12}. Outside this region anticipated synchronization becomes unstable and is lost abruptly. For values of the coupling strength $C$ too large or smaller than $0.5$ or a $\tau$ larger than 1 the value of $\rho$ significantly decreases. The optimal parameters for stable anticipated synchronization are in the range $C=0.5-1$ with the largest $\tau \sim 1$. So, in the next sections, we study the suppression of extreme events in the parameter region around $C=0.75$ and $\tau=1$.

\section{Supression of extreme events}
\label{prevention}

As previously discussed, the desynchronization events in the main system, also known as bubbling events, can be deterministic, due to small mismatches in the parameters, or stochastic, due to the presence of noise even when the master and the slave systems are identical \cite{Flunkert09,Tiana12}. In the following subsection we study the suppression of large deterministic desynchronization events induced by a parameter mismatch, while in subsection B we consider identical units under the influence of noise.

\subsection{Deterministic extreme events induced by parameter mismatch}
\label{deterministic}

In order to stabilize the synchronized state of the main system we consider a simple method of control based on corrective signals of an infinitesimal duration. This method was demonstrated to be effective in the suppression of spikes in coupled Adler \cite{Mayol12} and in two coupled FitzHugh-Nagumo \cite{Ciszak09} systems. It consists in applying a reset of fixed voltage amplitude $V_c \delta(t-t')$, that lasts only one integration time step ($dt= 1$ to $4 \times 10^{-4}$ in our case), when the variable $y_{\perp}$ reaches a predefined threshold, $Th_2$. This reset is applied to the variable $V_{S_1}$ after the threshold is reached as $V_{S_1}(t)\to V_{S_1}(t)+V_c$ if $V_{M_2}>V_{S_2}$ and $V_{S_1}(t)\to V_{S_1}(t)-V_c$ if $V_{M_2}<V_{S_2}$ in order to reduce $|V_{M_1}-V_{S_1}|$. So, we must know simultaneously the predicted variable $y_{\perp}$ and the sign of the difference $V_{M_2}-V_{S_2}$.

\begin{figure}[]
\includegraphics[width=8cm] {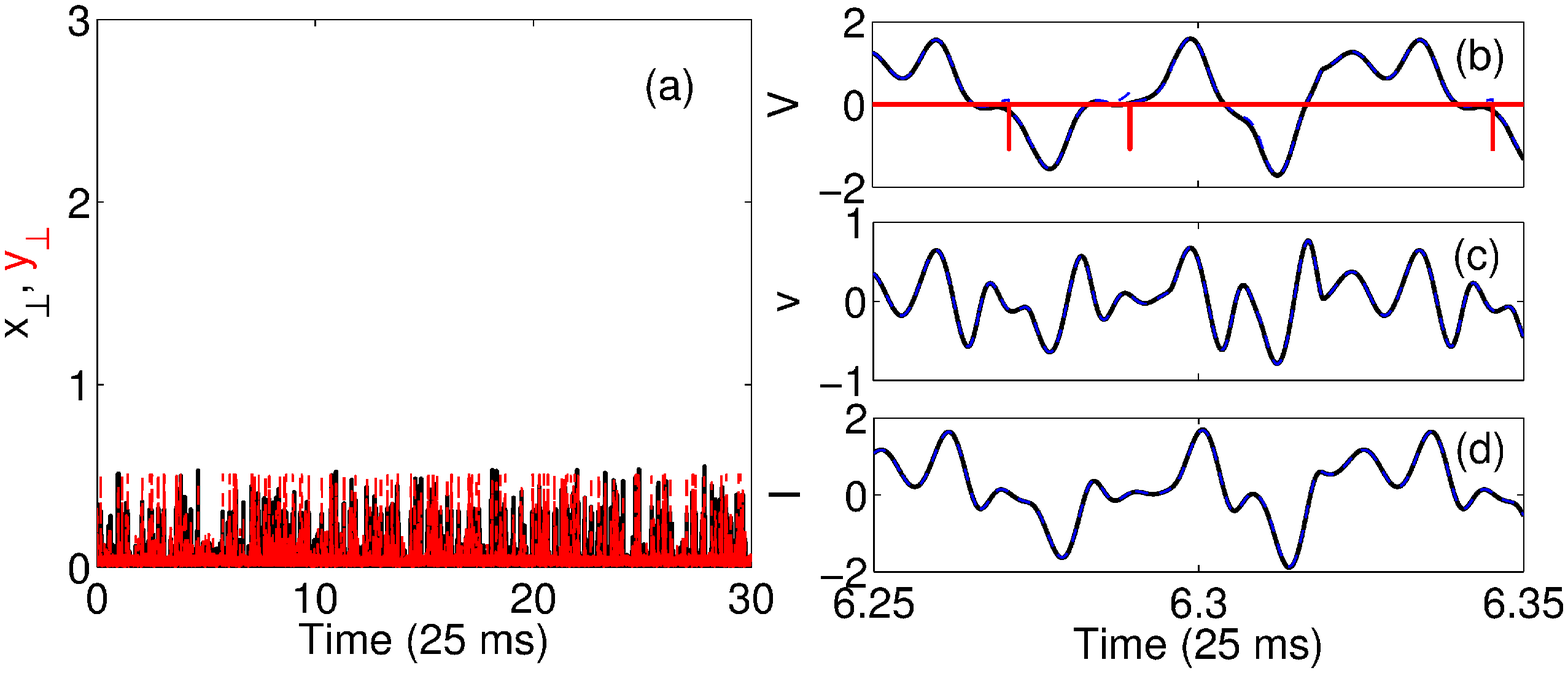}
\includegraphics[width=8cm] {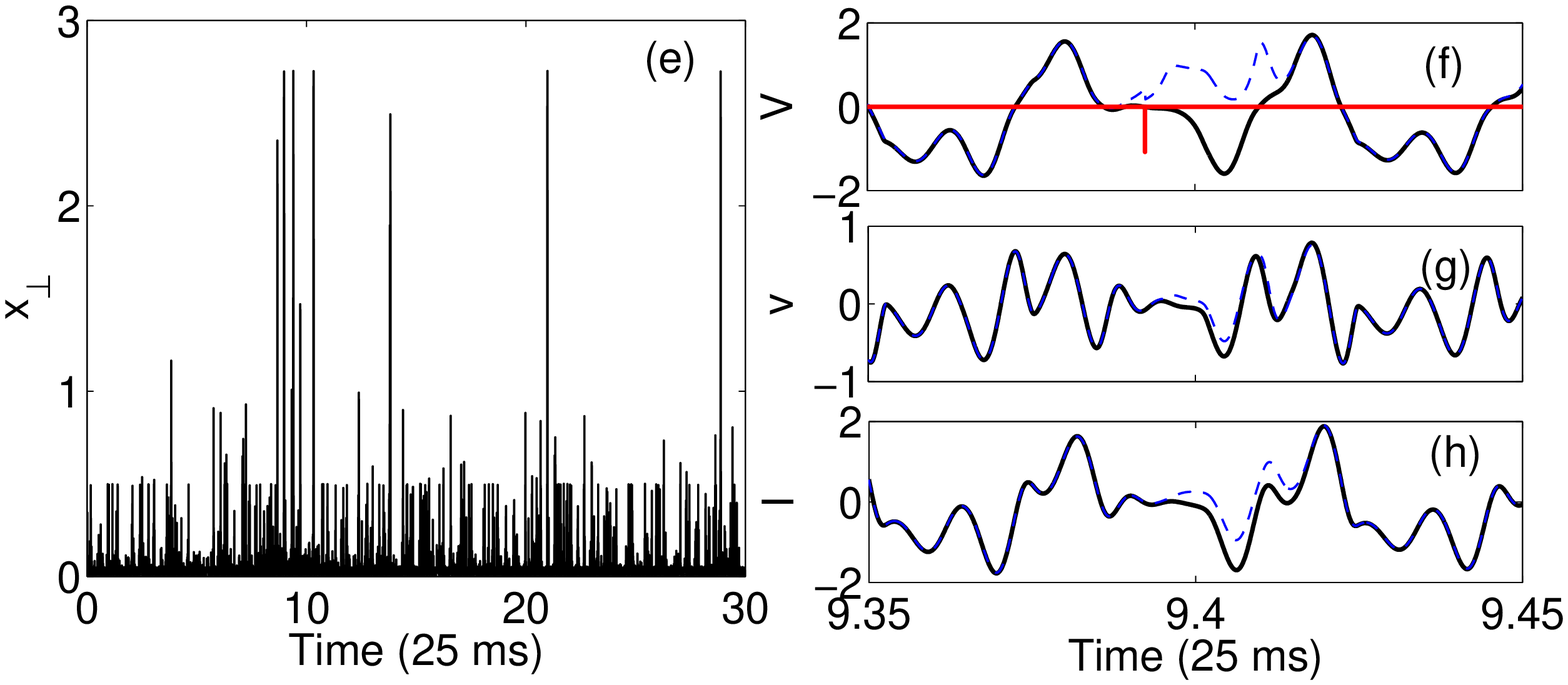}
\caption{\label{fig:5} (Color online) (a) Time trace for $x_{\perp}$ (black line) and $y_{\perp}$ (red line). The control is triggered when $y_{\perp}$ exceeds the threshold $Th_2=0.5$. (b)-(d) Time trace of each of the variables of the main system $M_1$ (black solid line) and $S_1$ (blue dashed line). In panel (b) we also show a magnification of the control (red line). (e) Time trace for $x_{\perp}$ setting that the control turns on when $x_{\perp}$ exceeds the threshold $Th_1=0.5$. (f)-(h) Time trace of each of the variables of the main system $M_1$ (black solid line) and $S_1$ (blue dashed line) checking $x_{\perp}$. In panel (f) we also show a magnification of the control (red line). $|V_c|=$ 0.27 and the integration time step is $dt=4\times 10^{-4}$.}
\end{figure}

A typical time trace is shown in figure \ref{fig:5} (a) where the threshold is fixed at $Th_2=0.5$. When the auxiliary system crosses this threshold the control is triggered and remains inactive 7 units of time after its application in order to avoid additional triggers in a single event. Before the control is triggered $x_{\perp}$ is smaller than $Th_2$ and rarely reaches or surpasses this value after the reset. As shown in figures \ref{fig:5} (b)-(d), $M_1$ and $S_1$ remain synchronized due to the control that efficiently brings $S_1$ to the evolution of $M_1$. Note that if we want to be more selective and act only on the larger events the corrective resets must be applied at larger threshold values, requiring large voltages $V_c$, i. e. more energy, to efficiently suppress the event. 

We can also apply the corrective resets directly on the main system. A time trace is shown in figure \ref{fig:5} (e) for the same set of parameters used in the previous case but now triggering the control when $x_{\perp}$ reaches the threshold ($Th_1=0.5$). In this case, $x_{\perp}$ is larger when the reset is triggered and the suppression becomes less effective unless we use larger corrective reset voltages. Afterwards the reset, $x_{\perp}$ decreases but the reduction is not large enough to bring $M_1$ and $S_1$ together and in some cases the difference keeps growing (figures \ref{fig:5} (f)-(h)) resulting in a poor control over the main system. By comparing figures \ref{fig:5} (a) and \ref{fig:5} (e) almost complete suppression of extreme events is obtained for a control through the predicted variable $y_{\perp}$ while for thresholds defined using $x_{\perp}$ the corrective reset is too small and some extreme events are still observed in the main system. Consequently, anticipated synchronization improves 
the performance of the control by applying corrective resets of small amplitude some time in advance.

\begin{figure}[]
\includegraphics[width=8cm] {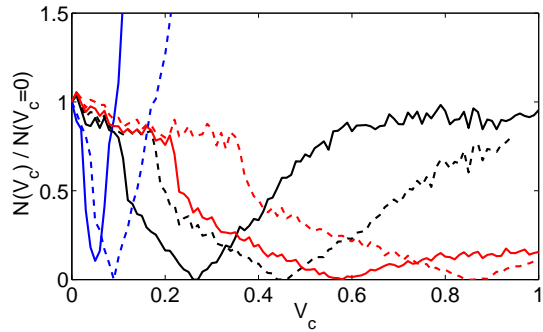}
\caption{\label{fig:6} (Color online) Fraction of detected DKs for a reset amplitude $V_c$ divided by the number of DKs when the control is off for threshold values $Th_1=0.1$ (blue dashed line), $Th_2=0.1$ (blue solid line), $Th_1=0.5$ (black dashed line), $Th_2=0.5$ (black solid line), $Th_1=1$ (red dashed line), and $Th_2=1$ (red solid line).}
\end{figure}

We have explored a broad range of parameter using even larger mismatches and in all cases the control through the auxiliary system was better. Here, a better control means that we can suppress the extreme events with currents $V_c$ a 30-40\% smaller applied during the same time. To show this, we now calculate the DKs that survive to the control. A Dragon King is precisely defined as a desynchronization event that increases above $x_{\perp}=2$ ($y_{\perp}=2$) and lasts until $x_{\perp}<0.5$ ($y_{\perp}<0.5$). In figure \ref{fig:6} we show the fraction of surviving DKs, defined as the number $N$ of DKs detected after the control divided by the number of DKs detected without control, $N(V_c=0)$, for different values of the threshold, $Th_{1,2}$. The voltage, $V_c$, is considered optimal when almost all DKs are suppressed and is a 30-40\% smaller when checking the predicted signal (solid lines) compared with the direct detection (dashed lines). Moreover, for the optimal value of $V_c$ checking the predicted 
signal 30-40\% of the DKs remain when using the direct detection. It is worth mentioning that above the optimal voltage the fraction of remaining DKs grows with $V_c$ because the large resets applied for the control change the sign of $V_{M_2}-V_{S_2}$. The selected threshold also affects the sensitivity of the suppression due to the "V'' shape around the optimal $V_c$. Large thresholds lead to broader ranges of high suppression (blue and red lines) while the number of surviving DKs is more sensitive to variations of $V_c$ for small thresholds (blue line).

\begin{figure}[]
\includegraphics[width=8cm] {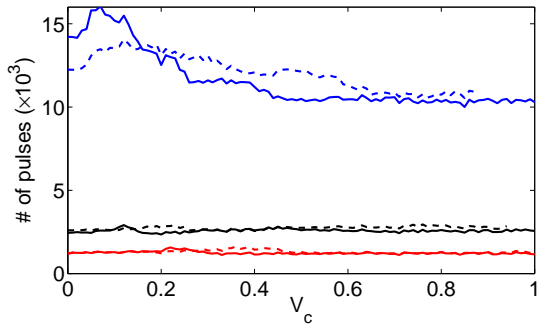}
\caption{\label{fig:7} (Color online) Number of control resets as a function of $V_c$ for threshold values $Th_1=0.1$ (blue dashed line), $Th_2=0.1$ (blue solid line), $Th_1=0.5$ (black dashed line), $Th_2=0.5$ (black solid line), $Th_1=1$ (red dashed line), and $Th_2=1$ (red solid line).}
\end{figure}

The number of corrective resets is obtained for realizations of a fixed total time ($T=8\times$10$^5$ units of time) and shown in fig. \ref{fig:7} as a function of $V_c$. This value is much larger for small thresholds $Th_{1,2}=0.1$ (blue lines) than for large thresholds (red and black lines) as expected from the power-law distribution of desynchronization events. Low thresholds have the inconvenience of triggering the control for a large number of events that do not lead to extreme events, while, at the same time, they are unable to suppress the extreme events for $V_c>0.1$ as shown in figure \ref{fig:6}. Larger thresholds, $Th_{1,2}=0.5$ and $Th_{1,2}=1$ lead to a small number of resets in the range of optimal control. Note also that for $V_c=0$ and $Th_{1,2}=0.1$ the number of resets is larger when checking $y_{\perp}$ (solid blue line) than when checking $x_{\perp}$ (dashed blue line) due to the larger parameter mismatch in the auxiliary system (see Table \ref{tab:T1}). Despite the fact 
that the results shown in figs. \ref{fig:6} to \ref{fig:8} do not depend on the integration time step ($dt= 10^{-4}$ in our case), for too large $V_c$ above the region of optimal $V_c$, the integration diverges for some parameters due to the sharpness of the resets (e. g. figure \ref{fig:6} black dashed line for $V_c>0.9$) and a smaller time step is required.

\subsection{Stochastic extreme events induced by noise}
\label{stochastic}

Bubbling events can also appear due to noise although $M_1$ and $S_1$ are identical. Desynchronization events can not be predicted just by knowing the current state of the system due to the unpredictability of noise. However, here we show that even in the case of DKs induced by noise, we are able to predict the appearance of such DKs with certain anticipation time and apply the control method explained in the previous section to prevent the extreme events.

Accordingly to the previous description, we set all the parameters identical to the parameters of $M_1$ listed in Table \ref{tab:T1} and introduce independent white noises of zero mean and delta correlated in time to each variable, $\zeta_i$, $\xi_i$, and $\eta_i$ but with the same intensity, $D$, i. e. the equations for the main system are modified from eqs. (\ref{eq:V1})-(\ref{eq:I1}) becoming:
\begin{eqnarray}
\dot{V}_{j_1}&=&\frac{V_{j_1}}{R_{1,M_1}}-g_{j_1}\left[V_{j_1}-v_{j_1}\right]+ D\zeta_{j_1}, \\
\dot{v}_{j_1}&=&g_{j_1}\left[V_{j_1}-v_{j_1}\right]-I_{j_1}+ D\xi_{j_1},\\
\dot{I}_{j_1}&=&v_{j_1}-R_{4,M_1}I_{j_1}+D\eta_{j_1},\end{eqnarray}
where
\begin{eqnarray}
g_{j}\left[V\right]&=&\frac{V}{R_{2,M_1}}+I_{r,M_1}\left(e^{\alpha_{f,M_1}V}-e^{-\alpha_{r,M_1}V}\right).\end{eqnarray}
Similarly, the auxiliary system is modified from eqs. (\ref{eq:AuxV})-(\ref{eq:AuxI}) and can be written as:
\begin{eqnarray}
\dot{V}_{j_2}&=&\frac{V_{j_2}}{R_{1,M_1}}-g_{j_2}\left[V_{j_2}-v_{j_2}\right]+C\left(V_{j_1}-V_{j_2,\tau} \right)\\ \nonumber
&&+ D\zeta_{j_2},\\
\dot{v}_{j_2}&=&g_{j_2}\left[V_{j_2}-v_{j_2}\right]-I_{j_2}+ D\xi_{j_2}\\
\dot{I}_{j_2}&=&v_{j_2}-R_{4,M_1}I_{j_2}+D\eta_{j_2},
\end{eqnarray}
where the coupling is introduced in both systems as was done in Section \ref{model}. For $D=0$, $M_1$ and $S_1$ are synchronized at all times, i. e. $x_{\perp}=0$. For $D> 0$ noise introduces small perturbations that in certain regions of the phase space can lead to divergent trajectories. The amplitude and frequency of the bubbling events depend on the size of the perturbation and the region of the phase space where noise acts.

\begin{figure}[]
\includegraphics[width=8cm] {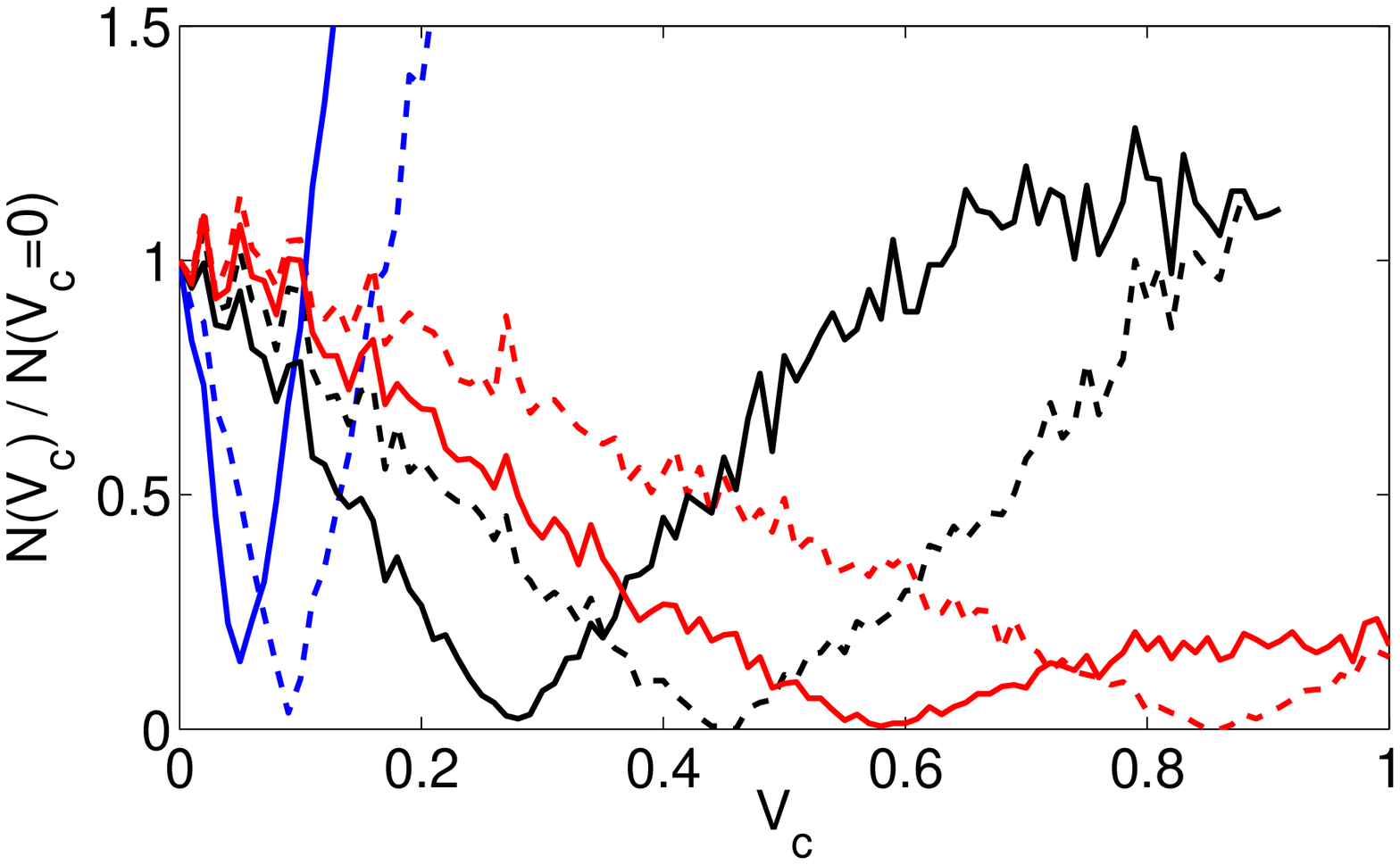}
\caption{\label{fig:8} (Color online) Fraction of detected DKs for a reset amplitude $V_c$ divided by the number of DKs when the control is off for threshold values $Th_1=0.1$ (blue dashed line), $Th_2=0.1$ (blue solid line), $Th_1=0.5$ (black dashed line), $Th_2=0.5$ (black solid line), $Th_1=1$ (red dashed line), and $Th_2=1$ (red solid line). $D=3\times10^{-3}$.}
\end{figure}

Now, we apply the same control as in the deterministic case. Interestingly, the ratio of surviving stochastic DKs as a function of $V_c$ is comparable to the deterministic case as shown in fig. \ref{fig:8} which suggests that noise plays a role similar to the parameters mismatch \cite{Tessone07}.

\begin{figure}[]
\includegraphics[width=8cm] {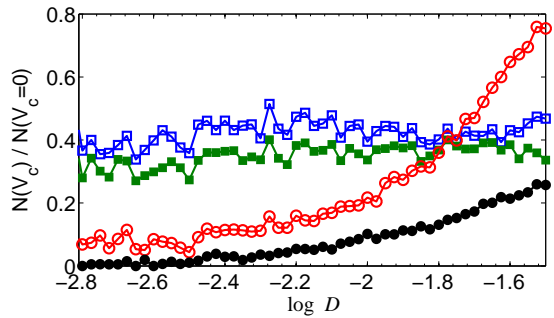}
\caption{\label{fig:9} (Color online) Fraction of detected DKs when the control is on divided by the number of DKs when the control is off as a function of the logarithm of the noise strength, $D$. $Th_1=0.2$, $V_c=0.1$ (open squares); $Th_1=0.5$, $V_c=0.28$ (filled squares); $Th_2=0.2$ and $V_c=0.1$ (open circles); and $Th_2=0.5$, $V_c=0.28$ (filled circles). $dt= 4\times 10^{-4}$.}
\end{figure}

The fraction of surviving DKs after the reset is shown in figure \ref{fig:9} as a function of the noise strength. If the threshold is checked in $x_{\perp}$ (open and filled squares) the noise strength has almost no effect on the fraction of DKs which remains around 40\%. However, when the threshold is checked in $y_{\perp}$ (open and filled circles) the fraction of surviving DKs can be substantially smaller. For small $D$'s the fraction is almost zero and monotonously increases when increasing $D$. This is consistent with the results obtained with different sets of parameters in the deterministic case. In figure \ref{fig:4} the accuracy of the prediction, as well as the anticipation time, depends on the set of parameters used in the system. When the parameters are very different the region of stable anticipated synchronization decreases and also does the anticipation time. In the stochastic case, we find that the region of anticipated synchronization decreases smoothly by increasing $D$. The 
quality of the prediction strongly affects the efficiency of the control using anticipated synchronization and suggests that the anticipation method would not be a good option when the extreme events are induced by a large noise.

\section{Conclusions}
\label{conclusions}

We have shown that anticipated synchronization can be used to predict the appearance of extreme desynchronization events known as Dragon Kings. We have shown that a corrective reset can be combined with anticipated synchronization to efficiently reduce and almost completely suppress these extreme events. In the case of deterministic DKs induced by parameter mismatches, the detection of desynchronization events in the predicted signal requires smaller amplitudes of the resets when compared with a direct detection in the main system in order to suppress the DKs. We have also shown that noise can induce DKs playing a role similar to the parameter mismatch. Also in this case, anticipated synchronization can be efficiently used to suppress noise-induced DKs if the noise strength is not too large.

We speculate that the results obtained here can be improved in some systems if the control is applied during the precursor of the extreme event \cite{Zamora13} leading to larger anticipation times.

\section{Acknowledgements}
\label{acknowledgements}
This work was supported by MINECO (Spain), Comunitat Aut\`onoma de les Illes Balears, FEDER, and the European Commission under the INTENSE@COSYP project FIS2012-30634


\begin{thebibliography}{99}
\bibitem{footnote} The delay time for which anticipated synchronization occurs, e. g. see figure \ref{fig:3}, is relatively small compared with the time scale of the extreme events and is smaller than in other systems. We speculate that this is due to the strong nonlinearities of the system and expect larger anticipation times in systems with weaker nonlinearities.
\end{thebibliography}
\end{document}